# Mitigation of Side-Effect Modulation in Optical OFDM VLC Systems

MOHAMMED M. A. MOHAMMED, CUIWEI HE, AND JEAN ARMSTRONG, (Fellow, IEEE)
Department of Electrical and Computer Systems Engineering, Monash University, Melbourne, VIC 3800, Australia
Corresponding author: Mohammed M. A. Mohammed (mohammed.ma.mohammed@monash.edu)

AQ:1 This work was supported by the Australian Research Council's (ARC) Discovery funding schemes under Grant DP 150100003 and Grant DP 180100872.

**ABSTRACT** Side-effect modulation (SEM) has the potential to be a significant source of interference in future visible light communication (VLC) systems. SEM is a variation in the intensity of the light emitted by a luminaire and is usually a side effect caused by the power supply used to drive the luminaires. For LED luminaires powered by a switched mode power supply, the SEM can be at much higher frequencies than that emitted by conventional incandescent or fluorescent lighting. It has been shown that the SEM caused by commercially available LED luminaires is often periodic and of low power. In this paper, we investigate the impact of typical forms of SEM on the performance of optical OFDM VLC systems; both ACO-OFDM and DCO-OFDM are considered. Our results show that even low levels of SEM power can significantly degrade the bit-error-rate performance. To solve this problem, an SEM mitigation scheme is described. The mitigation scheme is decision-directed and is based on estimating and subtracting the fundamental component of the SEM from the received signal. We describe two forms of the algorithm; one uses blind estimation, while the other uses pilot-assisted estimation based on a training sequence. Decision errors, resulting in decision noise, limit the performance of the blind estimator even when estimation is based on very long signals. However, the pilot system can achieve more accurate estimations, and thus a better performance. Results are first presented for typical SEM waveforms for the case where the fundamental frequency of the SEM is known. The algorithms are then extended to include a frequency estimation step and the mitigation algorithm is shown also to be effective in this case.

**INDEX TERMS** ACO-OFDM, DCO-OFDM, estimation, intensity modulated direct-detection OFDM, interference, VLC.

## I. INTRODUCTION

Light emitting diodes (LEDs) are increasingly being used as illumination devices for indoor and outdoor applications [1]. This widespread adoption of LEDs provides the opportunity to utilize them also for data communication. In practical applications of visible light communication (VLC), the visible light signal is vulnerable to noise and interference. There are two major sources of noise; the shot noise caused by ambient light and thermal noise caused by the receiver circuits [2].

Interference in VLC systems can be due to a number of different sources. While it is well-known that incandescent and fluorescent lights can cause interference particularly at low frequencies [3], [4], what is less well-known is that LED luminaires are also a potential source of interference. This interference may be at higher levels and at higher frequencies than interference caused by conventional light sources [5]–[7] and so has the potential to be a significant problem for VLC. The form of this interference is very dependent on how the LEDs are powered and, if the lights are dimmable, how dimming is achieved.

The time variation of light output from luminaires has long been a concern of those working in the lighting industry. In the past, various terms were used to describe this time variation including flicker, flutter, and shimmer. However, in more recent work, other terms including 'modulation' have been used [8]. In this paper, we term this modulation Side-Effect Modulation (SEM) to distinguish it from the intentional modulation of light in VLC which we simply term 'modulation'.

Studies have shown that SEM can have a wide range of biological effects including photosensitive epileptic seizures, headaches, reduced concentration, and phantom array effects [5], [6]. As a result, standards have been developed for acceptable SEM levels for frequencies up to 3 kHz [8]. Because there are no known biological effects for frequencies above 3 kHz, these standards set no limit on high-frequency SEM (HF-SEM), however, this form of interference can significantly degrade VLC performance.

In this paper, we study the impact of SEM on VLC systems and analyze its effect when optical orthogonal frequency division multiplexing (OFDM) is used as the modulation technique. OFDM is an advantageous technology for VLC because of its high spectral efficiency and immunity to inter-symbol interference (ISI) [9].

Several techniques have been proposed to overcome interference in VLC systems and these can also be used to mitigate SEM. One of these is Manchester coding, which has been shown to be effective, however, it reduces the achievable data rate [10]. Another technique exploits the fact that the interference is typically concentrated in the low-frequency region. Therefore, it can be reduced by using an electrical high pass filter at the receiver. However, a significant disadvantage of this approach is that it may introduce significant ISI [11]. In OFDM, the influence of interference can often be reduced by simply leaving the low-frequency subcarriers unused but this incurs some loss in spectral efficiency [12]. Unfortunately, this approach cannot be used for HF-SEM as the frequency depends on the properties of the interfering source which is generally unknown at the transmitter.

In VLC, intensity modulation with direct detection (IM/DD) is always used due to the non-coherent characteristics of LEDs [13]. As the intensity of the light is directly modulated, the transmitted signal must be real and unipolar. The most common unipolar OFDM techniques are DC biased OFDM (DCO-OFDM) [14] and asymmetrically clipped optical OFDM (ACO-OFDM) [15].

In this paper, we describe an SEM mitigation technique, in which the fundamental component of the SEM is estimated and then subtracted from the received signal. To the best of our knowledge, there is no research on mitigating SEM in optical OFDM, and this is the first paper to describe a digital signal processing (DSP) based SEM mitigation scheme. We show that this technique can substantially reduce the SEM without any loss in spectral efficiency. The effect of SEM is analyzed for both ACO-OFDM and DCO-OFDM.

The main contributions made in this paper are summarized as follows;
1) The effect of SEM on optical OFDM VLC is analyzed for both ACO-OFDM and DCO-OFDM.
2) The relationship between the amount of distortion introduced by SEM and the frequency of the SEM is described.
3) An SEM mitigation approach, which is based on SEM fundamental component estimation, is described. Two algorithms are designed. In the first, decision-directed estimation is employed and in the second, pilot-assisted estimation is used. The accuracy of each algorithm is evaluated.
4) Finally, the performance improvement achieved by the described technique is assessed for ACO-OFDM and DCO-OFDM.

The rest of the paper is organized as follows: in Section II, SEM is described and its impact on ACO-OFDM and DCO-OFDM is demonstrated. In Section III, the SEM mitigation technique is introduced. The BER simulation results before and after mitigation are presented and discussed in Section IV. Finally, the paper is concluded in Section V.

## II. IMPACT OF SEM IN VISIBLE LIGHT COMMUNICATION

In this section, we start by describing the SEM. Then the transmitter, the receiver, and the channel model for VLC in the presence of SEM are presented. Finally, we describe the effect of SEM on system performance.

### A. SIDE-EFFECT MODULATION (SEM)

SEM is likely to be an important impairment in many practical VLC applications. This is because often not all of the LEDs in a given scenario will be used for VLC. There may also be some illumination-only LEDs and possibly some legacy incandescent and fluorescent luminaires. These are all potential sources of SEM.

The spectral content of the SEM generated by LED luminaires depends very much on the details of the circuits which convert from the ac power source to the dc supply to the LEDs. There is often a strong component at twice the line frequency, which is either 100 Hz or 120 Hz depending on the country, but there may also be much higher frequencies caused by switched mode power supplies. Components between 2 to 150 KHz have been reported in the literature [16], [17] and measurements in our laboratory found that the LED luminaires in the room had components at 380 Hz and 62 KHz.

SEM can be divided into two main categories: low-frequency SEM (LF-SEM) and HF-SEM. While LF-SEM may have a serious impact on health [6], its effects in VLC can be minimized in a carefully designed system. It can be reduced by using a high pass filter at the receiver. In systems using OFDM, LF-SEM can be avoided by not using the first subcarrier for data transmission. On the other hand, while HF-SEM has no known health effects, we will show that even low levels of HF-SEM can significantly degrade the performance of VLC systems.

It has been shown that the SEM generated by LED luminaires is often periodic and can take various forms including sinusoidal waveforms, or square waveforms [7], [8]. SEM often consists of a fundamental frequency and harmonics of this frequency but typically, the fundamental component is much stronger than the harmonics.

### B. OFDM TRANSMITTER AND RECEIVER STRUCTURE

In this section we show the structure of a VLC system in the presence of SEM and when either ACO-OFDM or DCO-OFDM is used. In this paper, perfect synchronization is assumed between the transmitter and the receiver.

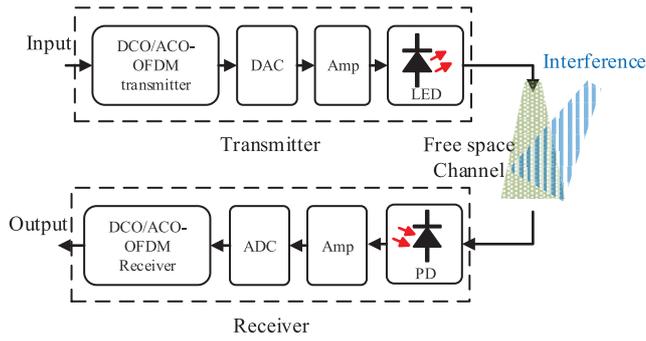

**FIGURE 1.** Visible light communication system block diagram.

Fig. 1 shows the VLC transmitter and receiver structure. After the ACO-OFDM or DCO-OFDM signal is generated, it is converted to an analog signal using a digital-to-analog converter (DAC). The analog signal is then amplified and used to directly modulate a LED. In ACO-OFDM, only odd subcarriers are used for modulation and for DCO-OFDM all subcarriers are used. Hence, for the same constellation size, DCO-OFDM can achieve higher data rates than ACO-OFDM [18].

The OFDM systems considered in this paper use a $N$ point FFT/IFFT. $N$ also represents the total number of subcarriers. After adding a cyclic prefix (CP) with length $N_{CP}$, the length of one OFDM symbol becomes $N_T = N + N_{CP}$. The frequency of the $k$-th subcarrier, for $0 \leq k \leq N/2 - 1$, is given by

$$f_k = kf_0, \quad (1)$$

where $f_0 = 2B/N$ and $B$ is the bandwidth.

At the receiver, a photodiode (PD) is used to detect the light signal and convert it back to an electrical signal. This signal contains the transmitted signal, the SEM from other light sources, and noise. Next, the signal is amplified and converted back to digital using an analog-to-digital (ADC) converter. Finally, the transmitted data symbols are recovered from the odd subcarriers if ACO-OFDM is used, or from all subcarriers, if DCO-OFDM is used.

In the presence of SEM, the received sampled signal can be expressed as

$$y(n) = x(n) + s_{\text{SEM}}(n) + w(n), \quad (2)$$

where $x(n)$ is the sampled transmitted ACO-OFDM or DCO-OFDM signal, $s_{\text{SEM}}(n)$ is the sampled SEM, and $w(n)$ is the sampled additive white Gaussian noise (AWGN), which models the shot and thermal noise.

### C. EFFECT OF SEM
In this subsection, the impact of SEM on ACO-OFDM and DCO-OFDM is studied. The optical power of the signal, $E\{x(n)\}$, is set to unity. For a fair comparison, the SEM variance, $E\{s_{\text{SEM}}(n)^2\}$, is fixed for all SEM waveforms. Sinusoidal, clipped sinusoidal, sawtooth, and square SEM waveforms are considered. We first show the impact on performance for a fixed frequency and a fixed SEM variance. Next, we show how the change in SEM variance affects the performance. Finally, we explain the relationship between the SEM frequency and the degradation in performance. In the following, we define $l$ as the normalized frequency so $l = f/f_0$ where $f$ is the absolute frequency.

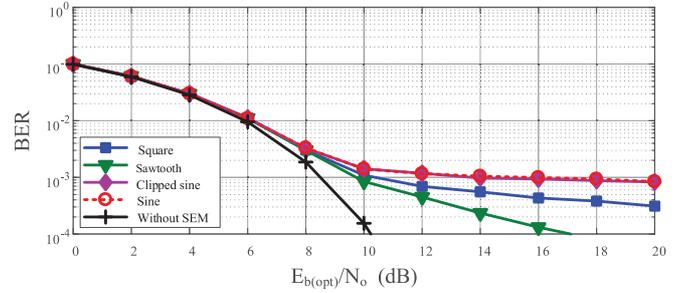

**FIGURE 2.** $E_{b(\text{opt})}/N_0$ versus BER for ACO-OFDM, over an AWGN channel using 16-QAM and different SEM waveforms and fixed SEM variance $\sigma_s^2 = 0.005$ with $l = 2.56$.

The bit error rate (BER) plots against the received[1] optical energy per bit to noise power ratio, $E_{b(\text{opt})}/N_0$, are used to evaluate the performance. In Fig. 2, we show the influence of SEM with $l = 2.56$, and a small variance of $\sigma_s^2 = 0.005$ on the BER performance of 16-QAM ACO-OFDM. Similarly, Fig. 3 (a) and (b) show the results for 16-QAM DCO-OFDM with 7 dB and 13 dB bias levels. The bias in dB is calculated as $10\log_{10}(1+\mu^2)$ dB where $\mu$ is set relative to the standard deviation of the unclipped signal [18]. In this paper, $l = 2.56$ corresponds to 10 kHz, which falls within the actual range of frequencies found in practice [16]. Also, the level of variance is chosen so that it falls within the practical levels [7], [8]. SEM degrades the BER performance in all three cases even when the SEM power is small. Different SEMs introduce different levels of degradation, this is because different SEM waveforms possess harmonics with different power levels. As a result, they affect the data subcarriers differently.

For the same optical power, the impact of SEM on DCO-OFDM is much greater than for ACO-OFDM. This is due to three factors; first, the large portion of power allocated to the DC bias, which reduces the power allocated to the data-carrying subcarriers compared with ACO-OFDM; second, the clipping distortion associated with DCO-OFDM; third, the nulled even subcarriers in ACO-OFDM. Note that the clipping distortion introduced by a 13 dB bias is smaller than a 7 dB [18]. For a fixed optical power, the allocated power on the data-carrying subcarriers decreases as the DC bias level increases.

---

[1]In this paper we normalize the received power rather than the transmitted power, as the signal and SEM will in general be from different sources and experience different channels. This is distinct from papers which analyze the effect of receiver position on performance, as the transmit power in VLC is usually fixed and the received power depends on the distance between transmitter and receiver.

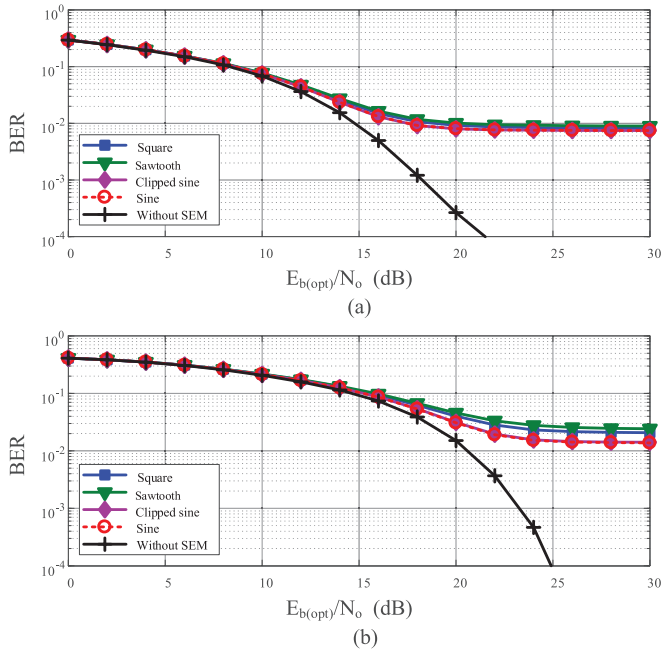

**FIGURE 3.** $E_{b(opt)}/N_0$ versus BER for DCO-OFDM, over an AWGN channel using 16-QAM and different SEM waveforms and fixed SEM variance $\sigma_s^2 = 0.005$ with $l = 2.56$, (a) 7dB bias level, (b) 13 dB bias level.

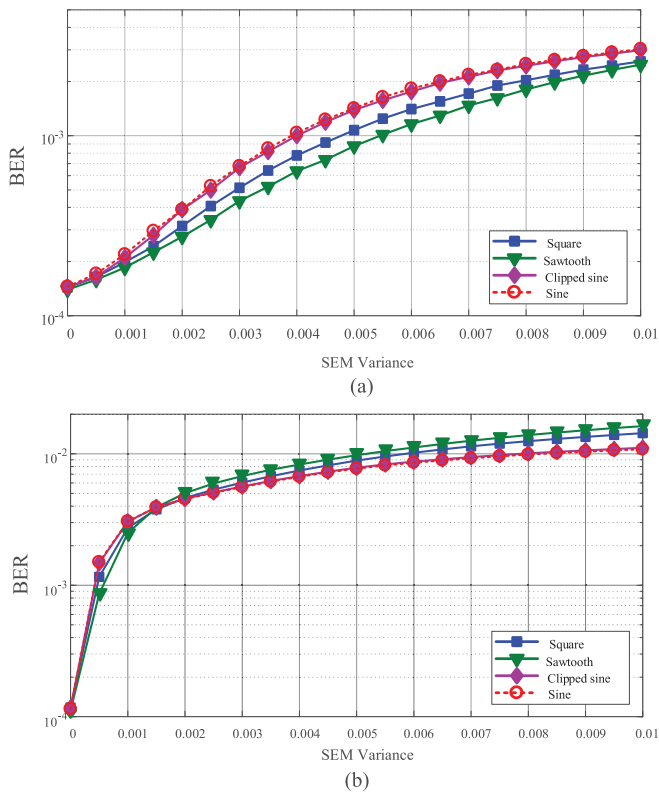

**FIGURE 4.** BER versus the SEM variance over an AWGN channel using 16-QAM and different SEM waveforms with $l = 2.56$ and fixed $E_{b(opt)}/N_0$, (a) ACO-OFDM and $E_{b(opt)}/N_0 = 10$ dB, (b) DCO-OFDM with 7 dB bias $E_{b(opt)}/N_0 = 21$ dB.

The effect of variable SEM power with $l = 2.56$ on performance for ACO-OFDM and DCO-OFDM with 7 dB bias is depicted in Fig. 4 (a) and (b). Four different waveforms are considered. The $E_{b(opt)}/N_0$ is fixed at 10 dB for ACO-OFDM and at 21 dB for DCO-OFDM. These values correspond to a low BER around $10^{-4}$ when there is no SEM. For ACO-OFDM, the sinusoidal waveform causes more errors and the sawtooth waveform causes fewer errors than other SEM waveforms. For DCO-OFDM and for SEM variance less than 0.0015, the sinusoidal SEM causes the most errors and the sawtooth SEM causes the fewest. However, for SEM variance greater than 0.0015, the sawtooth SEM causes the most errors and the sinusoidal causes the least. This is a result of the increase of the impact of the harmonics on performance as the variance increases. At low variances, the harmonics power is close to the noise floor so the degradation in performance is mainly because of the fundamental component. However, as the variance increases, the harmonics power increases and contribute more to the degradation in performance.

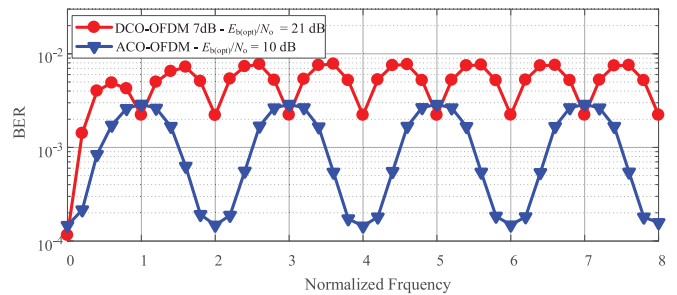

**FIGURE 5.** BER versus the SEM normalized frequency for ACO-OFDM and DCO-OFDM with 7 dB bias, over an AWGN channel using 16-QAM and sinusoidal SEM with $\sigma_s^2 = 0.005$ and fixed $E_{b(opt)}/N_0$.

The impact of changing the SEM frequency on the performance of ACO-OFDM and DCO-OFDM with 7 dB bias is shown in Fig. 5. A sinusoidal SEM with a fixed variance, $\sigma_s^2 = 0.005$, is used. $E_{b(opt)}/N_0$ is fixed at 10 dB for ACO-OFDM and at 21 dB for DCO-OFDM. In general, the impact of SEM on BER depends mainly on the fractional offset in frequency rather than the absolute frequency. The performance has an oscillating pattern between high and low BERs as the frequency changes. The impact of SEM also decreases when the frequency is close to the zeroth subcarrier. This is because the zeroth subcarrier does not carry data.

For DCO-OFDM, the impact of SEM peaks when the SEM frequency is midway between two subcarriers as in this case both subcarriers are affected by the SEM fundamental component. The impact has a trough when the frequency coincides with one of the data subcarriers as now only one subcarrier is affected.

For ACO-OFDM, the peaks are on the odd subcarriers and the troughs are on the even subcarriers. This is because the even subcarriers are not used in the receiver.

## III. SEM MITIGATION TECHNIQUE
In this section, we describe a way of mitigating SEM that does not result in any loss in spectral efficiency. This technique is based on estimating the SEM fundamental component and then subtracting it from the received signal. The advantage of this technique is that the receiver is not required to have any prior knowledge about the waveform type of the SEM. In the following, we start by describing a decision-directed SEM estimation algorithm in which no pilot data is transmitted, i.e. blind estimation, and demonstrate the performance improvement achieved by this technique. Next, we show how the estimation accuracy and performance can be improved if pilot-assisted estimation using a training sequence is used instead. We begin by considering the case where the frequency of the SEM fundamental component is known. Later, we discuss how the technique can be extended to the situation when the frequency is unknown.

## A. DECISION-DIRECTED ESTIMATION: ALGORITHM DESCRIPTION

In the absence of noise, the samples of the fundamental component of $s_{\text{SEM}}(n)$ are given by

$$s_F(n) = A\cos(\frac{2\pi nl}{N} + \theta_0), \quad n = 0, 1, ..., L \times N_T - 1, \quad (3)$$

where $A$ is the amplitude, $\theta_0$ is the initial phase, $l$ is the normalized frequency of the SEM fundamental component, and $L$ is the number of OFDM symbols. The initial phase, $\theta_0$, is the phase of the interfering signal at the start of the first OFDM symbol. It follows that the phase of SEM at the beginning of the $i$-th OFDM symbol is given by

$$\theta_i = \theta_0 + 2\pi l (1 + N_{\text{CP}}/N)(i - 1). \quad (4)$$

In the decision-directed (blind estimation) technique, the estimation is conducted in two stages. See Fig. 6. The steps in the first stage are identical to those in a conventional receiver. For clarity, in the following, we considered the case when no CP is added as this does not affect the performance of the algorithm.

In the first stage, the samples of the received signal, $y(n)$, are input to an FFT. The discrete frequency-domain FFT output of the OFDM symbol is given by

$$Y^{(i)}(k) = \frac{1}{\sqrt{N}} \sum_{n=0}^{N-1} y^{(i)}(n) \exp(\frac{-j2\pi nk}{N}). \quad (5)$$

After this stage, an initial decision is made on the received data using a maximum-likelihood estimator.

$$\hat{X}^{(i)}(k) = \arg\min_{X \in \Im_{M\text{-QAM}}} \left\| \alpha Y^{(i)}(k) - X \right\|,$$
$$= X^{(i)}(k) + E(k), \quad (6)$$

where $\alpha = 1$ for DCO-OFDM and $\alpha = 2$ for ACO-OFDM, $\Im_{M\text{-QAM}}$ is the $M$-QAM constellation space, and $E(k)$ is the decision noise.

In the second stage, the estimated data, $\hat{X}^{(i)}(k)$, are used to recreate an estimate of the transmitted signal using ACO-OFDM/DCO-OFDM modulator to get

$$\hat{\mathbf{x}}^{(i)} = \left[ \hat{x}^{(i)}(0), \hat{x}^{(i)}(1), \ldots, \hat{x}^{(i)}(N-1) \right]. \quad (7)$$

Here, we consider a signal with a sequence of $L$ OFDM symbols, which is given in vector form by

$$\hat{\mathbf{x}} = \left[ \hat{\mathbf{x}}^{(1)}, \hat{\mathbf{x}}^{(2)}, \ldots, \hat{\mathbf{x}}^{(L)} \right]. \quad (8)$$

This signal is then subtracted from the received signal,

$$\tilde{s}_{\text{SEM}}(n) = y(n) - \hat{x}(n),$$
$$= s_{\text{SEM}}(n) + w(n) - e(n). \quad (9)$$

The resultant signal, $\tilde{s}_{\text{SEM}}(n)$, in (9) contains the SEM and two sources of noise; the decision noise, $e(n)$ and the Gaussian noise $w(n)$. Next, the amplitude and the phase of the fundamental component, $s_F(n)$, are estimated. The estimation is performed using the well-known least squares (LS) estimator

$$\hat{A}, \hat{\theta}_0 = \arg\min_{\breve{A}, \breve{\theta}_0} \left\{ \sum_{n=0}^{N-1} \left( \tilde{s}_{\text{SEM}}(n) - \breve{s}_F(n) \right)^2 \right\}, \quad (10)$$

where

$$\breve{s}_F(n) = \breve{A}\cos\left(\frac{2\pi nl}{N} + \breve{\theta}_0\right), \quad (11)$$

is the SEM fundamental component to be estimated. $\hat{A}$ and $\hat{\theta}_0$ are the estimated amplitude and phase.

To simplify the minimization problem in (10), $\breve{s}_F(n)$ can be defined as

$$\breve{A}\cos(\frac{2\pi nl}{N} + \breve{\theta}_0) = \breve{A}\cos(\breve{\theta}_0)\cos(\frac{2\pi nl}{N})$$
$$- \breve{A}\sin(\breve{\theta}_0)\sin(\frac{2\pi nl}{N})$$
$$= \breve{a}_1\cos(\frac{2\pi nl}{N}) - \breve{a}_2\sin(\frac{2\pi nl}{N}), \quad (12)$$

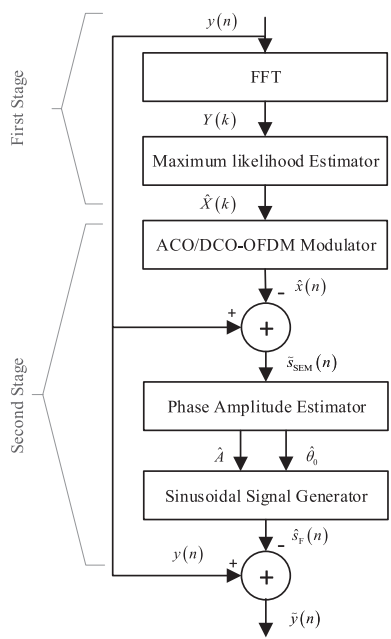

**FIGURE 6.** Decision-directed estimation algorithm.

where $\breve{a}_1 = \breve{A}\cos(\breve{\theta}_0)$ and $\breve{a}_2 = \breve{A}\sin(\breve{\theta}_0)$, so the new variables to be estimated are, $\breve{a}_1$ and $\breve{a}_2$. The LS estimator in (10) is expressed as

$$\hat{a}_1, \hat{a}_2 = \underset{\breve{a}_1, \breve{a}_2}{\arg\min}\left\{\sum_{n=0}^{N-1}\left(\tilde{s}_{\text{SEM}}(n) - \breve{a}_1\cos(\frac{2\pi nl}{N}) + \breve{a}_2\sin(\frac{2\pi nl}{N})\right)^2\right\}. \quad (13)$$

In vector form, (13) is given by

$$\hat{\mathbf{a}} = \arg\min_{\breve{\mathbf{a}}}\left\{\left(\tilde{\mathbf{s}}_{\text{SEM}} - \mathbf{G}\breve{\mathbf{a}}\right)^2\right\}, \quad (14)$$

where $\breve{\mathbf{a}} = [\breve{a}_1, \breve{a}_2]^T$, $\hat{\mathbf{a}} = [\hat{a}_1, \hat{a}_2]^T$, and

$$\mathbf{G} = \begin{bmatrix} 1 & 0 \\ \cos(\frac{2\pi l}{N}) & -\sin(\frac{2\pi l}{N}) \\ \vdots & \vdots \\ \cos(\frac{2\pi(N-1)l}{N}) & -\sin(\frac{2\pi(N-1)l}{N}) \end{bmatrix}. \quad (15)$$

Since $\mathbf{G}$ is known, the solution to this equation is given by

$$\hat{\mathbf{a}} = \left(\mathbf{G}^T\mathbf{G}\right)^{-1}\mathbf{G}^T\tilde{\mathbf{s}}_{\text{SEM}}. \quad (16)$$

The estimated phase and amplitude are given by

$$\hat{A} = \sqrt{\hat{a}_1^2 + \hat{a}_2^2}, \quad (17)$$

$$\hat{\theta}_0 = \arctan(\hat{a}_2/\hat{a}_1). \quad (18)$$

Next, the SEM fundamental component is recreated at the receiver using the estimated parameters and then subtracted from the received signal to give the signal with the mitigated SEM,

$$\tilde{y}(n) = y(n) - \hat{s}_F(n), \quad (19)$$

where

$$\hat{s}_F(n) = \hat{A}\cos(\frac{2\pi nl}{N} + \hat{\theta}_0). \quad (20)$$

Finally, $\tilde{y}(n)$ is sent to a conventional receiver for data detection. In the following sections, we evaluate the accuracy of the described algorithm in estimating the fundamental component of a SEM. Since we only estimate the fundamental component, in the following sections and without loss of generality a sinusoidal SEM is considered.

### B. DECISION-DIRECTED ESTIMATION: ESTIMATION ACCURACY

The effectiveness of the decision-directed estimation method is evaluated by calculating the root-mean-square (RMS) error for a variable number of symbols, $L$. For all cases, our simulation results showed that the RMS error in phase is small and it is always less than one degree. Therefore, in this section and in the following section, only in the results for amplitude are plotted.

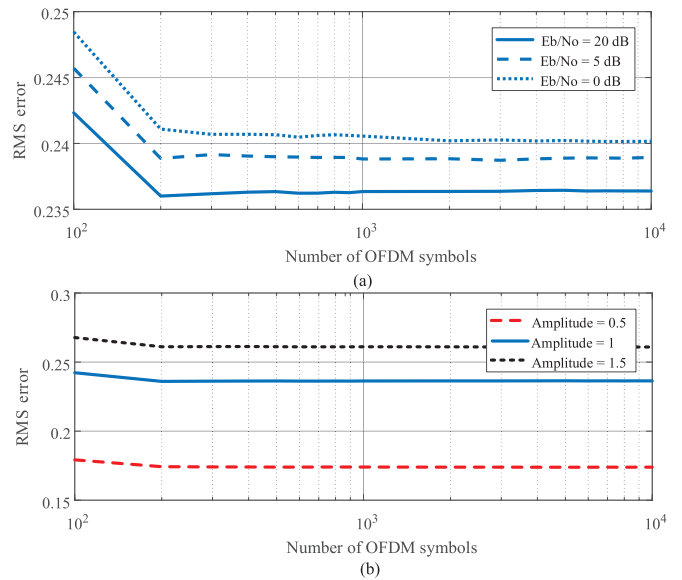

**FIGURE 7.** Amplitude RMS error in blind estimation for sinusoidal SEM with $l = 2.56$, for, (a) different $E_{b(\text{opt})}/N_0$; 0 dB, 5 dB, and 20 dB, and fixed amplitude $A = 1$ (b) different SEM levels; $A = 0.5$, $A = 1$, and $A = 1.5$, and fixed $E_{b(\text{opt})}/N_0 = 20$ dB.

In Fig. 7 (a), the RMS error in amplitude is plotted. Sinusoidal SEM with unity amplitude is considered and three different levels of $E_{b(\text{opt})}/N_0$ are used. The RMS errors in this figure and the following figures are calculated over $10^3$ estimates, $L$ is varied between $10^2$ and $10^4$ symbols, and 16-QAM ACO-OFDM is considered. It is clear that the error for the amplitude decreases as the number of OFDM symbols increases. The error also decreases as $E_{b(\text{opt})}/N_0$ increases. This is because the effect of the AWGN noise on the estimation decreases as $E_{b(\text{opt})}/N_0$ increases. For blind estimation, the RMS error in the amplitude does not converge to zero even if the number of symbols is large. This is due to the presence of the decision noise.

Fig. 7 (b) shows the RMS error in amplitude for $E_{b(\text{opt})}/N_0 = 20$ dB and different SEM levels. This figure shows the influence of the level of SEM on the accuracy of the estimation. The RMS error in estimating the amplitude increases as SEM level increases due to the increase in the decision noise.

### C. PILOT-ASSISTED ESTIMATION: ALGORITHM DESCRIPTION

As we saw in the previous section, decision-directed estimation can result in erroneous estimation because of the decision noise. An effective and practical way that can be used to improve the estimation is by using pilot data. This data is appended to the start of the transmitted OFDM signal and used during the channel estimation process to accurately estimate the SEM fundamental component. In this case, the error resulting from the data estimation process can be completely removed. This leads to a significant improvement in performance.

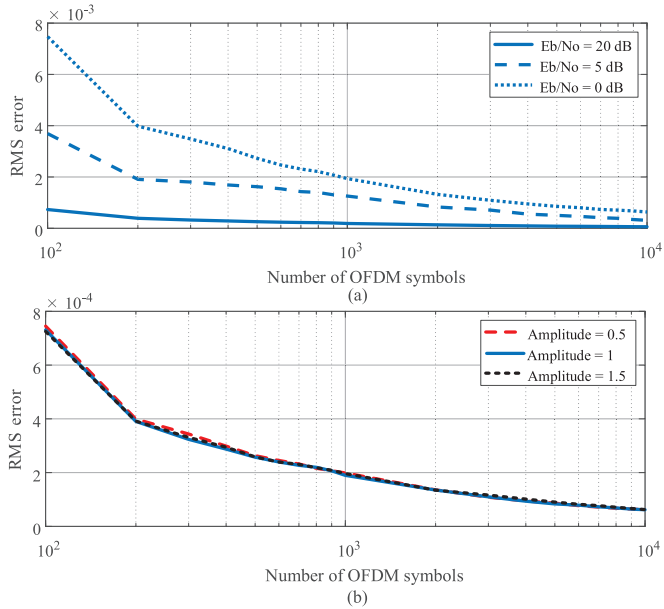

**FIGURE 8.** Amplitude RMS error in pilot-assisted estimation for sinusoidal SEM with $l = 2.56$, for, (a) different $E_{b(opt)}/N_0$ ; 0 dB, 5 dB, and 20 dB, and fixed amplitude $A = 1$ (b) different SEM levels; $A = 0.5$, $A = 1$, and $A = 1.5$, and fixed $E_{b(opt)}/N_0 = 20$ dB.

The pilot data symbols, $X_p(k)$, are used to create an ACO-OFDM or DCO-OFDM signal, $x_p(n)$. Note that $x_p(n)$ can alternatively be generated directly in the time-domain. This signal is then subtracted from the corresponding received signal, $y_p(n)$, to mitigate the SEM,

$$\tilde{s}_{SEM}(n) = y_p(n) - x_p(n). \quad (21)$$

After that, the rest of the algorithm is similar to the algorithm of blind estimation.

### D. PILOT-ASSISTED ESTIMATION: ESTIMATION ACCURACY

Fig. 8 (a) shows the results for pilot-assisted estimation for three different values of $E_{b(opt)}/N_0$. The dramatic improvement in the estimation accuracy achieved by pilot-assisted estimation compared with blind estimation can be seen by comparing Fig. 8 with Fig. 7. (Note the difference in scales for both axes). Pilot-assisted estimation is much more accurate and far fewer OFDM symbols are required in the estimation process. This means that pilot-assisted estimation not only improves the estimation but also substantially reduces the estimator complexity by reducing the amount of data involved in the estimation process.

Fig. 8 (b) shows how the accuracy of the pilot assisted estimator depends on the level of the SEM. It clearly shows that changing SEM level has very little impact on the RMS error in amplitude estimation. This is because there is no decision noise and in this case $E_{b(opt)}/N_0$ is high. The relatively low RMS errors shown in Fig. 8 prove the effectiveness of pilot-assisted estimator in reducing the decision noise and consequently improving the accuracy of the estimation.

## IV. BER SIMULATION RESULTS

We now demonstrate, using BER curves, the improvement that can be achieved by applying the mitigation technique. We show the performance improvement for different SEM waveforms.

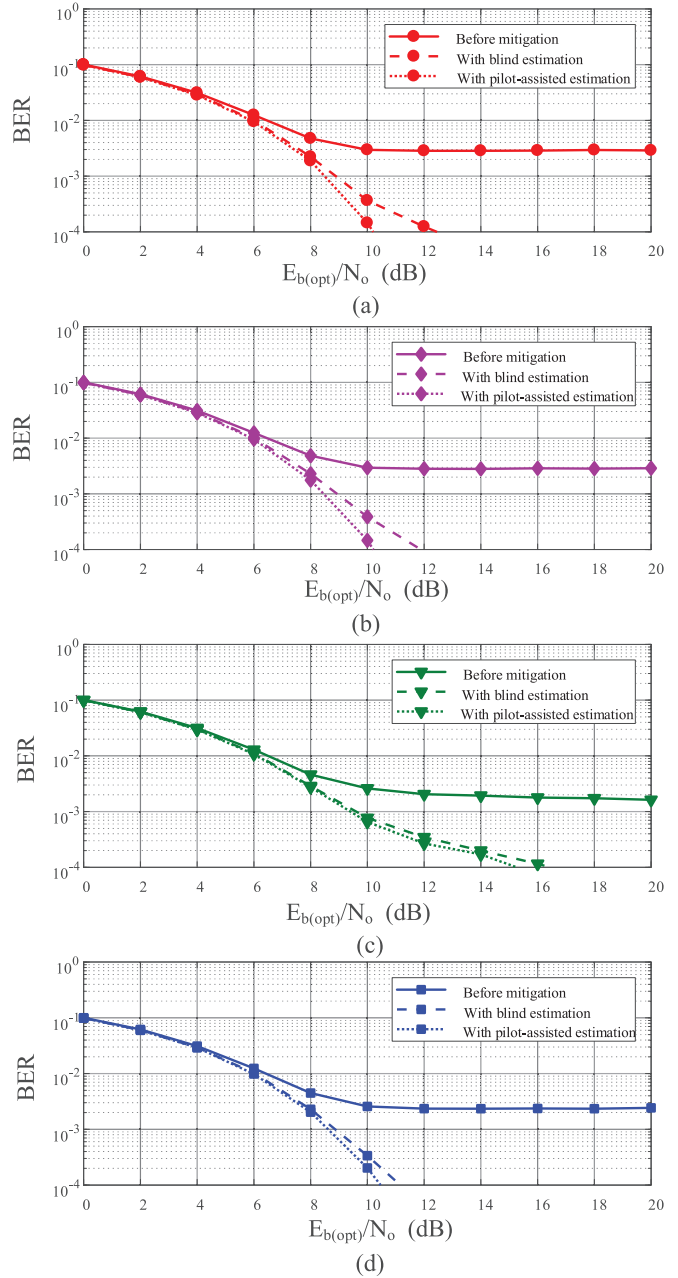

**FIGURE 9.** $E_{b(opt)}/N_0$ versus BER for 16-QAM ACO-OFDM, over an AWGN channel with $l = 2.56$ and $\sigma_s^2 = 0.01$ SEM using blind estimation, and pilot-assisted estimation, (a) sine, (b) clipped sine, (c) sawtooth, (d) square.

Fig. 9 and 10 show the BER results for ACO-OFDM and DCO-OFDM with 7 dB bias, respectively. These figures compare the BER before and after applying the mitigation techniques. Both blind and pilot assisted mitigation approaches are evaluated. Different SEM waveforms are used with $l = 2.56$, and $\sigma_s^2 = 0.01$. A sequence length of $L = 10^3$ is used for the estimation process.

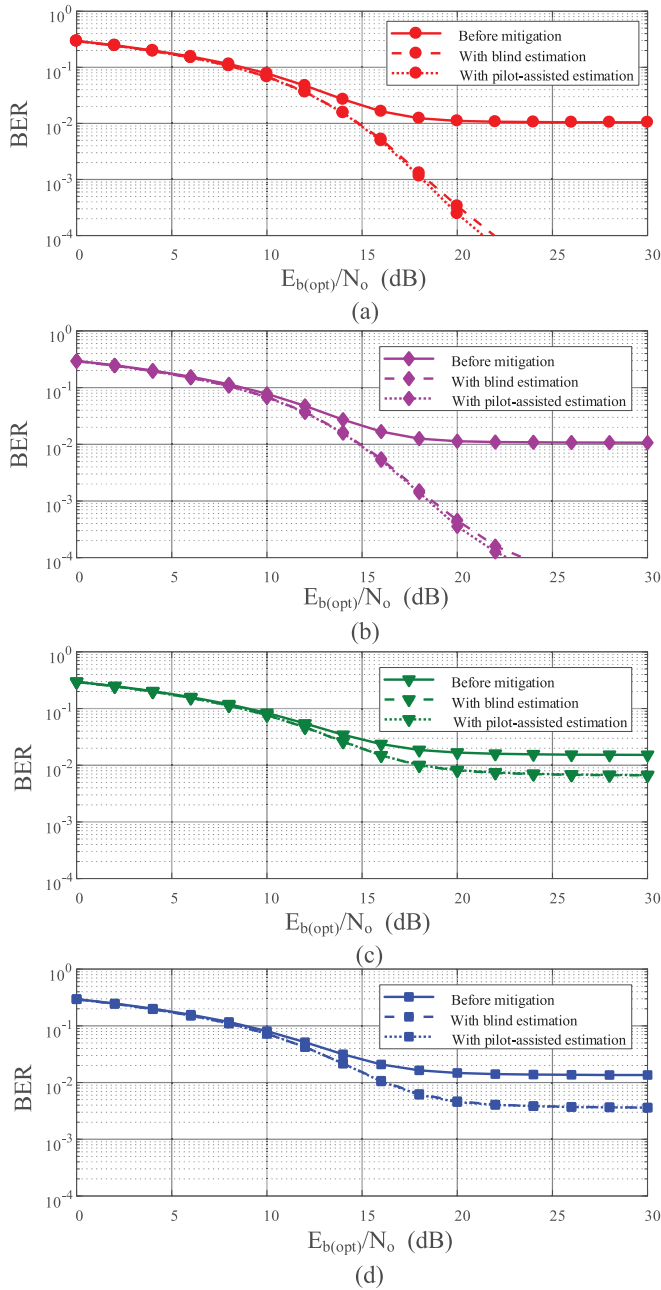

**FIGURE 10.** $E_{b(opt)}/N_0$ versus BER for 16-QAM DCO-OFDM with 7 dB bias, over an AWGN channel with $I = 2.56$ and $\sigma_s^2 = 0.01$ SEM using blind estimation, and pilot-assisted estimation, (a) sine, (b) clipped sine, (c) sawtooth, (d) square.

SEM mitigation using blind estimation significantly improves the performance. For sinusoidal SEM, the BER drops from $3 \times 10^{-3}$ to $4 \times 10^{-4}$ for ACO-OFDM at $E_{b(opt)}/N_0 = 10$ dB and from $1.3 \times 10^{-2}$ to $3.5 \times 10^{-4}$ for DCO-OFDM with 7 dB at $E_{b(opt)}/N_0 = 20$ dB. For a square SEM, the BER drops from $2.5 \times 10^{-3}$ to $3.5 \times 10^{-4}$ for ACO-OFDM and from $1.5 \times 10^{-2}$ to $4 \times 10^{-3}$ for DCO-OFDM. The improvement achieved by SEM mitigation is effective on all types of waveforms as it removes the fundamental component which contains most of the SEM power.

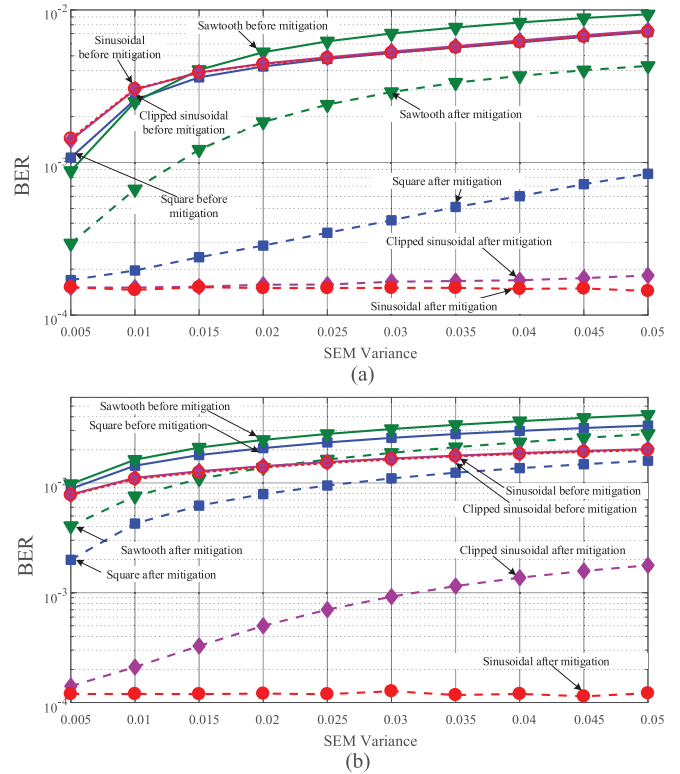

**FIGURE 11.** BER versus the SEM variance over an AWGN channel using 16-QAM and different SEM waveforms with $I = 2.56$ and fixed $E_{b(opt)}/N_0$, before and after pilot-assisted mitigation (a) ACO-OFDM and $E_{b(opt)}/N_0 = 10$ dB, (b) DCO-OFDM with 7 dB bias $E_{b(opt)}/N_0 = 21$ dB.

We now show that for the same sequence length of $L = 10^3$, pilot-assisted estimation results in lower BER than blind estimation for all cases. Fig. 11 shows the BER performance using pilot-assisted SEM mitigation on ACO-OFDM and DCO-OFDM with 7 dB bias. Variable SEM power and fixed $E_{b(opt)}/N_0$ are used. SEM mitigation is most effective for small SEM power regardless of the SEM waveform type.

In this paper, we consider four types of SEM waveforms. Although more complex waveform types can be found in practice, the described algorithm can be applied to remove the fundamental component and thus effectively improve the performance.

### A. EXTENSION TO UNKNOWN SEM FREQUENCY

In some practical cases, the frequency of the SEM fundamental component will be known. However, if the frequency is unknown the algorithm can be applied after an initial frequency estimation step. The frequency estimation problem is not a convex estimation problem [19] and there is no closed form solution, but the frequency can be estimated iteratively. An algorithm that can be used for finding the frequency is described in [20]. After the frequency is found, the amplitude and the phase are estimated as in the previous section.

Fig. 12 compares the estimation performance when the frequency is known and unknown for blind and pilot assisted estimations. It is shown that even if the frequency is unknown when the search algorithm described above is used to find the frequency, the performance for both blind and pilot-assisted estimation is very close to that achieved when the frequency is known. The estimation is very accurate when pilot assisted estimation is used.

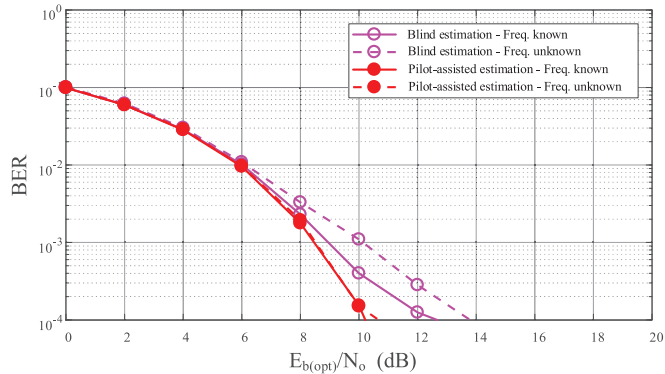

**FIGURE 12.** $E_{b(opt)}/N_0$ versus BER for 16-QAM ACO-OFDM over an AWGN channel with sinusoidal SEM with $I = 2.56$ and $\sigma_s^2 = 0.01$ in four cases: blind estimation with the frequency known and known, and pilot-assisted estimation with the frequency known and known.

## V. CONCLUSION

We have investigated the effect of SEM on the performance of an optical OFDM VLC system and shown that it can significantly degrade the performance for both DCO-OFDM and ACO-OFDM. A technique that can mitigate the effect of SEM has been described. This uses an LS estimator to estimate the SEM fundamental component which is then subtracted from the received signal. Two forms of the algorithm have been described, one using blind estimation and the second using a known pilot sequence. Results have been presented for the RMS error in amplitude. Decision errors, resulting in decision noise, limit the performance of the blind estimator even when estimation is based on very long sequences. The performance of the pilot system is better. The BER performance of the technique has been compared with systems using no mitigation for the case of an AWGN channel and 16-QAM modulation. The results show that the estimation technique significantly improves the performance of both ACO-OFDM and DCO-OFDM.

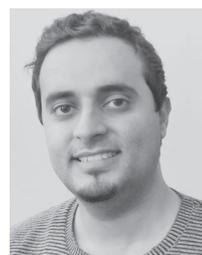

**MOHAMMED M. A. MOHAMMED** was born in Yemen. He received the B.Sc. degree in communication engineering from Damascus University, Damascus, Syria, in 2011, and the M.Sc. degree in optical communication from Alexandria University, Alexandria, Egypt, in 2015. He is currently pursuing the Ph.D. degree with the Department of Electrical and Computer Systems Engineering, Monash University, Melbourne, Australia. His main research interests are in optical wireless communication with a particular focus on intensity modulated direct-detection optical OFDM techniques.

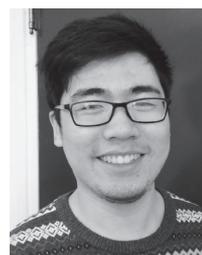

**CUIWEI HE** received the B.E. degree in telecommunication engineering from the Central South University of Forestry and Technology, Changsha, China, in 2010, the M.E. degree in digital signal processing from the University of the Ryukyus, Okinawa, Japan, in 2013, and the Ph.D. degree in optical wireless communication from Monash University, Melbourne, Australia, in 2017. He is currently a Research Fellow with the Department of Electrical and Computer Systems Engineering, Monash University. His research interests are digital signal processing, optical wireless communication, multiple-input multiple-output technology, and optical orthogonal frequency division multiplexing.

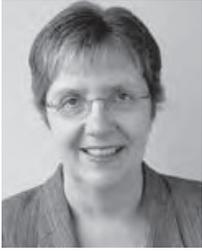

**JEAN ARMSTRONG** (M'90–SM'06–F'15) was born in Scotland. She received the B.Sc. degree (Hons.) in electrical engineering from The University of Edinburgh, Edinburgh, U.K., and the M.Sc. degree in digital techniques from Heriot-Watt University, Edinburgh. She is currently a Professor at Monash University, Melbourne, VIC, Australia, where she leads a research group working on topics, including optical wireless, radio frequency, and optical fiber communications. Before emigrating to Australia, she was a Design Engineer at Hewlett-Packard Ltd., Scotland. In Australia, she has held a range of academic positions at Monash University, the University of Melbourne, and La Trobe University. She has published numerous papers, including over 70 on the aspects of OFDM for wireless and optical communications and has six commercialized patents.

Mr. Armstrong has received numerous awards, including the Carolyn Haslett Memorial Scholarship, the Zonta Amelia Earhart Fellowship, the Peter Doherty Prize for the best commercialization opportunity in Australia (joint winner), induction into the Victorian Honour Roll of Women, the 2014 IEEE Communications Society Best Tutorial Paper Award, and the 2016 IET Mountbatten Medal. She is a fellow of the IEAust. She has served on the Australian Research Council (ARC) College of Experts and the ARC Research Evaluation Committee.

● ● ●